\documentclass[aps,prstab,preprint,showpacs]{revtex4}
\input{epsf}

\usepackage{graphicx}% Include figure files
\usepackage{dcolumn}% Align table columns on decimal point
\usepackage{bm}% bold math

\begin{document}

%\preprint{APS/123-QED}

\title{Two-mode entanglement in two-component Bose-Einstein
condensates}
\author{H. T. Ng and P. T. Leung\footnote{Email:
ptleung@phy.cuhk.edu.hk}}
\affiliation{{Department of Physics,
The Chinese University of Hong Kong,}\\
{Shatin, Hong Kong SAR, China}}
\date{\today}

\begin{abstract}
We study the generation of two-mode entanglement in a
two-component Bose-Einstein condensate trapped in a double-well
potential. By applying the Holstein-Primakoff transformation, we
show that the problem is exactly solvable as long as the number of
excitations due to atom-atom interactions remains low. In
particular, the condensate constitutes a symmetric Gaussian
system, thereby enabling its entanglement of formation to be
measured directly by the fluctuations in the quadratures of the
two constituent components [Giedke {\it et al.}, Phys. Rev. Lett.
{\bf 91}, 107901 (2003)]. We discover that significant two-mode
squeezing occurs in the condensate if the interspecies
interaction is sufficiently strong, which leads to strong
entanglement between the two components.
\end{abstract}

\def\aL{ \hat{a}^{\dag}_{L} }
\def\aR{ \hat{a}^{\dag}_{R} }
\def\bL{ \hat{b}^{\dag}_{L} }
\def\bR{ \hat{b}^{\dag}_{R} }
\def\Na{N_{\alpha}}
\def\Nb{N_{\beta}}
\def\wa{\Omega_{\alpha}}
\def\wb{\Omega_{\beta}}
\def\ka{\kappa_{\alpha}}
\def\kb{\kappa_{\beta}}
\def\kab{\kappa_{\alpha\beta}}

\pacs{03.75.Gg, 03.75.Lm, 03.75.Mn,03.67.Mn}
\maketitle

\section{introduction}
Soon after the experimental realization of Bose-Einstein
condensates (BECs), rich physical phenomena have been observed and
predicted as well \cite{expt,review,Leggett}. In particular, there
has been a surge of interest in the quantum tunneling dynamics of
BECs trapped in multiple wells, and much attention has been
focused on Josephson effect in such systems
\cite{review,Leggett,tunnel}. On the other hand, it was shown that
multi-particle entanglement can be generated in BECs via the
coherent interactions between the atoms \cite{multi}, and
thereafter BECs have played a prominent role in the field of
quantum information. For example, both multi-particle
entanglement (i.e. spin squeezing) and two-mode entanglement can
be generated in a spin-1 condensate with three hyperfine
sublevels \cite{duan,You}. In this case, two-mode entanglement,
describing the inseparability between the two modes (respectively
with spin projection $m=\pm 1$), can be used as quantum
information protocols to facilitate quantum teleportation of
continuous variables \cite{Zeilinger,Popescu}.

On the other hand, in addition to BECs with spin 0 and 1, BECs
with two internal degrees of freedom, e.g. the $|F=1,m=-1\rangle$
and $|F=2,m=2\rangle$ sublevels of ${}^{87}$Rb, are also
achievable experimentally and are often termed as two-component
condensates \cite{rubidium,Hall},  which give rise to novel
features such as phase separation \cite{Ho} and the cancellation
of mean field energy shift \cite{Meystre}. Such condensates can
be viewed as collections of interacting spin-half particles and
can consequently display multi-particle entanglement through the
emergence of spin squeezing \cite{Law}. Likewise, two different
kinds of atoms (e.g. ${}^{41}$K and ${}^{87}$Rb) can also form
stable two-component BECs \cite{KRbBEC}. More interestingly, the
interspecies interaction of these two-component BECs can be
varied by the application of magnetic control \cite{magint}, thus
hastening various experimental and theoretical investigations in
this field.

Recently, there have been several discussions on the dynamics of a
two-component condensate trapped in a double-well potential
\cite{lobo,Ng}. The behaviors of such systems, which can be
realized experimentally with the current technology, are arguably
much richer than those of single-component condensates because of
the interspecies interaction.   For example, in a recent paper
\cite{Ng} we studied the tunneling dynamics of a two-component
condensate whose two components are initially separated by the
potential barrier between a double-well. We found that in the
strong scattering regime atoms in the two components can tunnel
through the barrier in a correlated manner \cite{Ng}. As both the
scattering strength and the tunneling strength between the wells
  can be tuned independently with various
experimental techniques \cite{magint,Greiner}, it is expected that
such phenomena will become observable in the near future.

In this paper, we consider a two-component condensate trapped in a
double-well potential. Both components of the condensate are
initially prepared in the ground state of the double well and
hence are separable at time $t=0$. The objective of the present
paper is to study and quantify the generation of two-mode
entanglement in such a condensate. Two-mode entanglement is
commonly attributed to the inseparability of the density matrix
describing two systems, which can be found in various experimental
situations and is very useful in the applications of quantum
measurement and quantum information \cite{Zoller,Simon,Polzik}. In
fact, entanglement between two ensembles of atoms with spin has
recently been achieved experimentally via interaction with
polarized light \cite{Polzik}, and the concept of two-mode
entanglement has been generalized to describe such spin systems
\cite{duan}.  Since each component of a two-component condensate
in a double-well can be described as a collection of spin-half
particles, with the two spatially localized modes in the two wells
playing respectively the roles of spin-up and spin-down states,
the condensate is in fact equivalent to two interacting gigantic
spins that can be described in terms of continuous variables
\cite{duan}. While it has been shown that the intra-species
interaction is able to create spin-squeezing for a
single-component condensate trapped a double-well potential
\cite{Law}, in this paper we will demonstrate that the
inter-species interaction is  responsible for the generation of
two-mode entanglement. It is the intriguing interplay of the
inter-species and the intra-species interactions that sparks our
investigation in such systems. By applying the Holstein-Primakoff
transformation (HPT) \cite{HP}, which reduces this two-spin system
to two coupled oscillators, we show that our system is exactly
solvable as long as the number of excitations due to atom-atom
interactions remains low. More remarkably, the condensate in fact
forms a symmetric Gaussian system. Therefore, we can directly
evaluate the entanglement of formation (i.e. the von-Neumann
entropy for a pure state) of the system from the fluctuations in
the quadratures of the two constituent components
\cite{duan,Zoller,Simon,Giedke}. In accordance with the schemes
proposed recently by Duan {\it et al.} \cite{Zoller,Simon} and
Giedke {\it et al.} \cite{Giedke}, we define a two-mode
entanglement parameter that measures the fluctuations in the
quadratures and in turn analyze the degree of entanglement between
the two components of the condensate in the present paper. Our
discovery is that strong interspecies interaction can lead to
significant two-mode squeezing in the condensate and hence strong
entanglement between the two components.

The structure of our paper is as follows. In section II, we
introduce the Hamiltonian of our system and classify it into
symmetric and asymmetric cases according to the properties of the
constituent condensates. In section III, we consider specifically
a system with a large number of atoms.  An effective Hamiltonian,
which is exactly solvable, is then derived from the HPT. In
section IV, we introduce the two-mode entanglement parameter
\cite{duan,Zoller,Simon,Giedke} to describe and quantify the
entanglement in our system. In section V, we study in detail the
two-mode entanglement parameter for several typical cases.
Finally, we discuss the physical meaning of the two-mode
entanglement parameter in section VI and consider generalization
our approach to mixed states.

\section{Tunneling two-component BEC}
We first consider the tunneling dynamics of a two-component BEC
trapped in a symmetric double-well potential. The total number of
atoms in components $A$ and $B$ of the condensate are $N_a$ and
$N_b$ respectively. We will further assume that the interaction
 between the atoms is sufficiently weak and adopt the
two-mode approximation to describe the tunneling process. Under
such approximation, the condensate dwelling in each potential
minimum is adequately described by a single localized mode
function \cite{walls,raghavan,juha}. Considering the effect of
quantum tunneling and the conservation of the particle number of
each component, we obtain the Hamiltonian of the system:
\begin{eqnarray}
H &=& \frac{\Omega_a}{2} ({\aL}\hat{a}_{R}+{\aR}\hat{a}_{L})+
\frac{\Omega_b}{2}({\bL}\hat{b}_{R}+{\bR}\hat{b}_{L})
+{\kappa} ({\aL}\hat{a}_{L}{\bL}\hat{b}_{L}+{\aR}\hat{a}_{R}{\bR}\hat{b}_{R})
\nonumber\\ && +\frac{\kappa_{a}}{2}\left[
({\aL}\hat{a}_{L})^{2}+({\aR}\hat{a}_{R})^{2}\right]+
\frac{\kappa_{b}}{2}\left[({\bL}\hat{b}_{L})^{2}+({\bR}\hat{b}_{R})^{2}\right].
\label{Hamiltonian}
\end{eqnarray}
Here $\hat{a}^{\dag}_j$ ($\hat{a}^{}_j$) and $\hat{b}^{\dag}_j$
($\hat{b}^{}_j$) are respectively the creation (annihilation)
operators of components $A$ and $B$ residing in the $j$-th well,
$j=L,R$. Since there are two spatial modes (the $L$ and $R$
modes) available for each component, the Hamiltonian above in
fact consists of four bosonic operators. Besides, the parameters
$\Omega_a$($\Omega_b$), $\kappa_a$($\kappa_b$) and $\kappa$ are
the tunneling, intraspecies interaction strength of component
$A$($B$) and the interspecies interaction strength respectively.

For the convenience of the subsequent discussion of two-mode
entanglement, it is instructive to represent this Hamiltonian in
terms of angular momentum operators by following through the
standard Schwinger oscillator model to construct a set of spin
operators for each component \cite{Schwinger}:
\begin{eqnarray}
\hat{J}_{\alpha{x}}&=&\frac{1}{2}({\hat{\alpha}^{\dag}_{L}}\hat{\alpha}_{L}-
{\hat{\alpha}^{\dag}_R}\hat{\alpha}_{R}),\nonumber\\
\hat{J}_{\alpha{y}}&=&\frac{1}{2i}({\hat{\alpha}^{\dag}_{L}}\hat{\alpha}_{R}-
{\hat{\alpha}^{\dag}_R}\hat{\alpha}_{L}),\\
\hat{J}_{\alpha{z}}&=&\frac{1}{2}({\hat{\alpha}^{\dag}_{L}}\hat{\alpha}_{R}+
{\hat{\alpha}^{\dag}_R}\hat{\alpha}_{L}),
\nonumber
\end{eqnarray}
where $\alpha=a,b$. Here ${\hat{\bf
J}}_{\alpha}=(\hat{J}_{\alpha{x}},\hat{J}_{\alpha{y}},
\hat{J}_{\alpha{z}})$ obey the usual angular momentum commutation
relations. In the following we will  denote the eigenstates of
$\hat{J}^2_{\alpha} \equiv
\hat{J}_{\alpha{x}}^2+\hat{J}_{\alpha{y}}^2+
\hat{J}_{\alpha{z}}^2$ and $\hat{J}_{\alpha z}$ with
$|j_{\alpha},m_{\alpha}\rangle$ such that
$\hat{J}^2_{\alpha}|j_{\alpha},m_{\alpha}\rangle=j_{\alpha}(j_{\alpha}+1)
|j_{\alpha},m_{\alpha}\rangle$, and
$\hat{J}_{\alpha{z}}|j_{\alpha},m_{\alpha}\rangle=m_{\alpha}|j_{\alpha},m_{\alpha}\rangle$,
where $j_{\alpha}=N_{\alpha}/2$. In terms of $\hat{{\bf J}}_a$ and
$\hat{{\bf J}}_b$, the Hamiltonian (\ref{Hamiltonian}) can be
rewritten as
\begin{eqnarray}
H &=& {\Omega_a}\hat{J}_{az}+{\Omega_b}\hat{J}_{bz}+\kappa_{a}
\hat{J}^{2}_{ax} +\kappa_{b} \hat{J}^{2}_{bx}+
2{\kappa}\hat{J}_{ax}\hat{J}_{bx}.
\label{JHamiltonian}
\end{eqnarray}

%\subsection{Initial condition}
In the absence of atom-atom interactions, the ground state of the
system is obviously given by the product state
$|j_{a},-j_{a}{\rangle_{}}|j_{b},-j_{b}{\rangle_{}}$ and the two
components are not entangled. In the following discussion we
consider how the atom-atom interactions affect the evolution of
the initial state
\begin{equation}
|\Psi(t=0)\rangle=|j_{a},-j_{a}{\rangle_{}}|j_{b},-j_{b}{\rangle_{}}
\, ,\label{instate}
\end{equation}
and show that the two components will get entangled through the
inter- and intra-species interactions.

To facilitate later discussion on the phenomenon of entanglement,
we further classify two-component condensates according to the
symmetry properties of the two components constituting the
condensate. In a {\it symmetric} two-component BEC, the parameters
of component $A$ and component $B$ are equal to one another,
namely $\Omega_a=\Omega_b$ and $\kappa_a=\kappa_b$. These
conditions hold approximately for condensates consisting of the
hyperfine states $|F=1,m=-1\rangle$ and $|F=2,m=1\rangle$ of
${}^{87}$Rb \cite{Hall}. The two components of such condensates
have essentially same masses and magnetic moments and hence
$\Omega_a=\Omega_b$. Their intra-species scattering lengths are
quite close and, in addition,
$\kappa_a\approx\kappa_b\approx\kappa$ \cite{rubidium}. In a {\it
quasi-identical}  two-component BEC where
$\Omega_a=\Omega_b=\Omega$ and $\kappa_a=\kappa_b=\kappa$, the
Hamiltonian (\ref{JHamiltonian}) reduces to:
\begin{equation}\label{symconH}
H_{1}={\Omega}\hat{J_z}+{\kappa}\hat{J^2_x}.
\end{equation}
where $\hat{{\bf J}}=\hat{{\bf J}}_a+\hat{{\bf J}}_b$ is the total
angular momentum of the system. Despite that the Hamiltonian of
such a two-component condensate  is identical to that of a
single-component one \cite{Law}, the distinguishability of the two
species entails the study of two-mode entanglement.

Meanwhile, for {\it asymmetric} two-component BECs relevant
physical parameters of the two components are generally different.
For example, it has recently been observed in the experiment that
the condensates of potassium and rubidium (Rb-K), which have
different scattering lengths and masses, can form stable
two-component condensates \cite{KRbBEC}. Therefore, it is deemed
appropriate to develop a generic analytical scheme to study such
condensates. In the following discussion, we will make use of the
HPT \cite{HP} to carry out a thorough analytical investigation on
the entanglement between the two constituent components.

\section{Bosonic operator approximation }
In this section, we consider the evolution of a condensate with
large numbers of atoms and sufficiently weak scattering strengths,
namely $N_{a(b)}\gg 1$ and $\Omega_{a(b)}
{\gg}\kappa_a,\kappa_b,\kappa$. As the initial state, given by
(\ref{instate}), is the ground state of a non-interacting
condensate and the scattering strengths are weak, only the
low-lying states will be excited in the evolution and the
coherence of tunneling process can be maintained. The current
situation is in contradistinction to our previous study \cite{Ng}
that discovered correlated tunneling of the two components in the
strong scattering regime. However, we will show that the
inter-species interaction does lead to nontrivial entanglement of
the two components.

To proceed, we apply the HPT to map angular momentum operators
into bosonic operators \cite{HP,Bosonization} and show that under
the HPT our system is in fact equivalent to two coupled harmonic
oscillators. In HPT, the angular momentum operators
\begin{equation}
\hat{J}_{\alpha \pm}=\hat{J}_{\alpha x}\pm i\hat{J}_{\alpha
y}\,,\quad \alpha=a,b\,,
\end{equation}
and $\hat{J}_{\alpha{z}}$  are expressed in terms of bosonic
operators $\hat{\alpha}$, $\hat{\alpha}^{\dag}$:
\begin{eqnarray}
\hat{J}_{\alpha{+}}&=&\hat{\alpha}^{\dag}\sqrt{2j_\alpha
-\hat{\alpha}^{\dag}\hat{\alpha}},\\
\hat{J}_{\alpha{-}}&=&\left(\sqrt{2j_\alpha
-\hat{\alpha}^{\dag}\hat{\alpha}}\right)\hat{\alpha},\\
\hat{J}_{\alpha{z}}&=&(\hat{\alpha}^{\dag}\hat{\alpha}-j_\alpha).
\end{eqnarray}
Here $\hat{\alpha}^{\dag}$ and $\hat{\alpha}$ are standard
bosonic operators satisfying
$[\hat{\alpha},\hat{\alpha}^{\dag}]=1$. Hence, the Hamiltonian
(\ref{JHamiltonian}) can be written as
\begin{eqnarray}
H&=&\sum_{\alpha=a,b}\left[\Omega_\alpha({\hat{\alpha}^\dag}\hat{\alpha}
-j_\alpha)
+ \frac{\kappa_\alpha{j_\alpha}}{2}\Bigg(\hat{\alpha}^{\dag}\sqrt{1-
\frac{\hat{\alpha}^{\dag}\hat{\alpha}}{2j_\alpha}}
+\sqrt{1-\frac{\hat{\alpha}^{\dag}\hat{\alpha}}
{2j_\alpha}}\hat{\alpha}\Bigg)^2\right]
\nonumber\\
&&+{\kappa\sqrt{j_aj_b}}
\Bigg(\hat{a}^{\dag}\sqrt{1-\frac{\hat{a}^{\dag}\hat{a}}{2j_a}}
+\sqrt{1-\frac{\hat{a}^{\dag}\hat{a}}{2j_a}}\hat{a}\Bigg)
\Bigg(\hat{b}^{\dag}\sqrt{1-\frac{\hat{b}^{\dag}\hat{b}}{2j_b}}
+\sqrt{1-\frac{\hat{b}^{\dag}\hat{b}}{2j_b}}\hat{b}\Bigg).
\end{eqnarray}
Since $\Omega_{a(b)} \gg \kappa,\kappa_{a},\kappa_{b}$, it is
arguable that
\begin{equation}
\frac{\langle{\hat{\alpha}^{\dag}\hat{\alpha}}\rangle}{2j_\alpha} \ll 1,
\label{HPassumption}
\end{equation}
leading to an approximate effective Hamiltonian,
\begin{equation}
H_{\rm eff}=\Omega_a{\hat{a}^\dag{\hat{a}}}+\Omega_b{\hat{b}^\dag{\hat{b}}}
+\frac{1}{2}\Big[\kappa_aj_a(\hat{a}^{\dag}+\hat{a})^2
+\kappa_bj_b(\hat{b}^{\dag}+\hat{b})^2+
2\kappa\sqrt{j_aj_b}(\hat{a}^{\dag}+\hat{a})(\hat{b}^{\dag}+\hat{b})\Big].
\end{equation}
This Hamiltonian is analogous to that of two coupled oscillators
and completely captures the essence of the dynamics of the two
interacting BECs. Correspondingly, the initial state
(\ref{instate}) is given by the vacuum state $|0_a,0_b\rangle$ of
the two decoupled oscillators described by the first two terms in
$H_{\rm eff}$, where $|n_a,n_b\rangle$ represents the Fock state
of the oscillators.

To study the two-mode entanglement in the condensate, it is
advantageous to make use of the position and the momentum
operators:
\begin{eqnarray}
\hat{q}_a&=&\frac{1}{\sqrt{2}}(\hat{a}^{\dag}+\hat{a}),~~~~
\hat{p}_a=i\frac{1}{\sqrt{2}}(\hat{a}^{\dag}-\hat{a}),\\
\hat{q}_b&=&\frac{1}{\sqrt{2}}(\hat{b}^{\dag}+\hat{b}),~~~~
\hat{p}_b=i\frac{1}{\sqrt{2}}(\hat{b}^{\dag}-\hat{b}),
\end{eqnarray}
and to rewrite the effective Hamiltonian as
\begin{equation}
H_{\rm eff}=\frac{\Omega_a}{2}(\hat{q}_a^2+\hat{p}^2_a)
+\frac{\Omega_b}{2}(\hat{q}_b^2+\hat{p}^2_b) +\kappa_aj_a
\hat{q}_a^2+\kappa_bj_b\hat{q}_b^2+
2\kappa\sqrt{j_aj_b}\hat{q}_a\hat{q}_b.
\end{equation}
It is then straightforward to solve the resulting equations of
motion of $\hat{q}_a(t),\hat{p}_a(t),\hat{q}_b(t)$ and
$\hat{p}_b(t)$. For convenience, we express the solutions in
matrix form, which reads
\begin{eqnarray}
X(t)&\equiv&(\hat{q}_a(t),\hat{p}_a(t),\hat{q}_b(t),\hat{p}_b(t))^T \nonumber \\
&=&U(t)X(t=0).
\end{eqnarray}
Here $U(t)$ is a real  $4\times{4}$ matrix representing the
evolution operator and can be written as
\begin{displaymath}
U= \left( \begin{array}{clcr}
C & E_1  \\
E_2 & D
\end{array} \right),
\end{displaymath}
with the $2\times{2}$ matrices $C$, $D$ and $E$ being explicitly
given  by
\begin{displaymath}
C= \frac{1}{(\mu_2-\mu_1)}\left( \begin{array}{cc}
\mu_2\cos\omega_1{t}-\mu_1\cos\omega_2{t} &
\Omega_a(\mu_2\sin\omega_1{t}/\omega_1-\mu_1\sin\omega_2{t}/\omega_2) \\
-(\omega_1\mu_2\sin\omega_1{t}-\omega_2\mu_1\sin\omega_2{t})/\Omega_a &
\mu_2\cos\omega_1{t}-\mu_1\cos\omega_2{t}
\end{array} \right),
\end{displaymath}
\begin{displaymath}
D= \frac{1}{(\mu_2-\mu_1)}\left( \begin{array}{cc}
-\mu_1\cos\omega_1{t}+\mu_2\cos\omega_2{t} &
-\Omega_b(\mu_1\sin\omega_1{t}/\omega_1-\mu_2\sin\omega_2{t}/\omega_2)\\
(\mu_1\omega_1\sin\omega_1{t}-\mu_2\omega_2\sin\omega_2{t})/\Omega_b &
-\mu_1\cos\omega_1{t}+\mu_2\cos\omega_2{t}
\end{array} \right),
\end{displaymath}
\begin{displaymath}
E_1= \frac{1}{(\mu_2-\mu_1)}\left( \begin{array}{cc}
-(\cos\omega_1{t}-\cos\omega_2{t}) &
-\Omega_b(\sin\omega_1{t}/\omega_1
-\sin\omega_2{t}/\omega_2) \\
(\omega_1\sin\omega_1{t}-\omega_2\sin\omega_2{t})/\Omega_a &
-\Omega_b(\cos\omega_1{t}-\cos\omega_2{t})/\Omega_a
\end{array} \right),
\end{displaymath}
and
\begin{displaymath}
 E_2= \frac{1}{(\mu_2-\mu_1)}\left( \begin{array}{cc}
-\Omega_b(\cos\omega_1{t}-\cos\omega_2{t})/\Omega_a &
-\Omega_b(\sin\omega_1{t}/\omega_1
-\sin\omega_2{t}/\omega_2) \\
(\omega_1\sin\omega_1{t}-\omega_2\sin\omega_2{t})/\Omega_a &
-(\cos\omega_1{t}-\cos\omega_2{t})
\end{array} \right),
\end{displaymath}
where
\begin{eqnarray} \label{nmf2}
\mu_{1(2)}&=&\frac{\omega^2_{1(2)}-\Omega^2_a-2\kappa_aj_a\Omega_a}
{2\kappa\Omega_a\sqrt{j_aj_b}},
\end{eqnarray}
and the normal mode frequencies of the coupled oscillation are
\begin{eqnarray} \label{nmf}
\omega_{1(2)}&=&\Bigg\{\left(\frac{\Omega^2_a+\Omega^2_b}{2}
+\kappa_aj_a\Omega_a+\kappa_bj_b\Omega_b\right)\nonumber\\
&&\pm\left[\left(\frac{\Omega^2_a-\Omega^2_b
}{2}+\kappa_aj_a\Omega_a -\kappa_bj_b\Omega_b\right)^2
+4\kappa^2j_aj_b\Omega_a\Omega_b\right]^{1/2}\Bigg\}^{1/2}.
\end{eqnarray}

This approximate solution, which is based on HPT, is valid as long
as the condition (\ref{HPassumption}) holds. However, if the normal
frequency is complex, the system will become unstable. We can
therefore determine the stability condition from (\ref{nmf}):
\begin{eqnarray} \label{qsc}
|\kappa|<\frac{1}{2}\sqrt{\Bigg(\frac{\Omega_a}{j_a}+2\kappa_a\Bigg)
\Bigg(\frac{\Omega_b}{j_b}+2\kappa_b\Bigg)} \equiv \kappa_c.
\end{eqnarray}
the HPT fails to yield a self-consistent solution for systems
violating the inequality. It is interesting to note that in the
limit where $j_a,j_b\to\infty$, the condition for stability
reduces to:
\begin{eqnarray}\label{extend}
|\kappa|<\sqrt{\kappa_a\kappa_b} \equiv \kappa_e,
\end{eqnarray}
which is a well known result for BECs in extended space, and
violation of (\ref{extend}) will lead to the phase separation of
two component BECs \cite{Ho,phase-sep}. It is also worthwhile to
note that the stability criterion (\ref{qsc}) depends on the
numbers of atoms in the double-well and similar dependence has
previously been found for two-component BECs in a single well
\cite{phase-sep}. We will, however, assume the stability criterion
(\ref{qsc}) is satisfied throughout the present study and obtain
analytically the two-mode entanglement parameter for the
condensate, which will be defined in the following section. We
will see that in addition to yielding the analytic solution to the
tunneling dynamics, the HPT performed here also facilitates our
study on two-mode entanglement.

\section{Theory of Two-mode Entanglement}
Entanglement between two systems that are described in terms of
continuous variables is usually indicated by an inequality in its
Einstein-Podolsky-Rosen (EPR) uncertainty
\cite{duan,Zoller,Simon}:
\begin{eqnarray}
\label{EPRinequal}
\frac{1}{2}\left\{\langle[{\Delta}(\hat{q}_{a}+\hat{q}_{b})]^2\rangle+
\langle[{\Delta}(\hat{p}_{a}-\hat{p}_{b})]^2\rangle\right\}&<&1,
\end{eqnarray}
where $[\hat{q}_m,\hat{p}_n]=i\delta_{mn}$ for $m,n=a,b$.
$\hat{q}_{a(b)}$ and $\hat{p}_{a(b)}$ are respectively the
position and momentum operators (or any pairs of quadratures) of
system $a(b)$, and the above inequality simply implies that the
positions (momenta) of the particles are strongly anti-correlated
(correlated). In general, condition (\ref{EPRinequal}) is only a
sufficient condition for entanglement and does not imply
separability of the two systems even if it is violated
\cite{duan,Zoller}. However, it has recently been shown that a
necessary and sufficient condition for entanglement, which is
analogous to (\ref{EPRinequal}), can be established   if the
combined system is a Gaussian one in the sense that if its Wigner
characteristic function, defined by:
\begin{eqnarray}
\chi^{(w)}(\lambda_a,\lambda_b)&=&\text{tr}[{\rho}\text{exp}
({\lambda_a}\hat{a}-{\lambda^*_a}\hat{a}^{\dag}+{\lambda_b}\hat{b}
-{\lambda^*_b}\hat{b}^{\dag})] \,,
\end{eqnarray}
is a Gaussian function of $\lambda_a$ and $\lambda_b$
\cite{Zoller,Giedke}. Without loss of generality, one can assume
that the expectation values of all quadratures vanish and hence
the Wigner characteristic function of a Gaussian system is
expressible as:
\begin{eqnarray}
\chi^{(w)}(\lambda_a,\lambda_b) &=&\text{exp}\Bigg[-\frac{1}{2}
\Lambda^T M \Lambda \Bigg]\,\,,
\end{eqnarray}
where $M$ is a $4\times4$ real symmetric matrix and the matrix
$\Lambda$ is defined by
$\Lambda\equiv(\lambda^I_a,\lambda^R_a,\lambda^I_b,\lambda^R_b)^T$.
As the characteristic function can also be written as:
\begin{eqnarray}
\chi^{(w)}(\lambda_a,\lambda_b) &=&\text{tr}[{\rho}\text{exp}
(i\sqrt{2}\Lambda^T X)]\,,
\end{eqnarray}
it is obvious that the matrix elements of $M$ are  the correlation
functions of the quadrature variables
$X=(\hat{q}_a,\hat{p}_a,\hat{q}_b,\hat{p}_b)^T$. In fact,
$M_{ij}=\langle (X_i X_j +X_j X_i)\rangle  $ and therefore $M$ is
termed the covariance matrix.

As the amount of entanglement between the two systems is
unaffected by local unitary operations, say local rotations in
the $q$-$p$ plane and local squeezing operations, the matrix $M$
can be transformed into a standard form $M_s$ by several local
operations \cite{Zoller}:
\begin{displaymath}
M_s= \left( \begin{array}{clcr}
n_1 & 0 &  c_1 & 0 \\
0   &n_2&   0  & c_2 \\
c_1 & 0 &  m_1 & 0 \\
 0  &c_2&   0  & m_2
\end{array} \right),
\end{displaymath}
where $n_1,n_2,m_1,m_2$ are positive numbers, and
\begin{eqnarray}
\frac{n_1-1}{m_1-1}&=&\frac{n_2-1}{m_2-1},\\
|c_1|-|c_2|&=&\sqrt{(n_1-1)(m_1-1)}-\sqrt{(n_2-1)(m_2-1)}.
\end{eqnarray}
It has recently been shown by Duan {\it et al.} that a Gaussian
system is entangled if and only if the following inequality is
satisfied \cite{Zoller,Giedke}:
\begin{eqnarray}
{a^2_0(n_1+n_2)-2(|c_1|+|c_2|)+(m_1+m_2)/{a^2_0}}
<{2a^2_0+2/{a^2_0}},
\end{eqnarray}
where
\begin{eqnarray}
a^2_0&=&\sqrt{\frac{m_1-1}{n_1-1}}=\sqrt{\frac{m_2-1}{n_2-1}}\,.
\end{eqnarray}
We therefore accordingly construct a two-mode entanglement
parameter $\xi_t$\cite{Zoller,Giedke}:
\begin{eqnarray}
\xi_t\equiv\frac{{a^2_0(n_1+n_2)-2(|c_1|+|c_2|)+(m_1+m_2)/{a^2_0}}}{{2a^2_0+2/{a^2_0}}}\,.
\end{eqnarray}
Physically speaking, $\xi_t>0$ is merely a suitably weighted
EPR-type uncertainty in the squeezed quadratures of the system.
The sufficient and condition for entanglement mentioned above can
then be expressed in terms of an inequality involving this
parameter:
\begin{eqnarray}
\xi_t<1.
\end{eqnarray}

If, in addition, the system is a symmetric one such that
$n_1=n_2=m_1= m_2$, Giedke {\it et al.} \cite{Giedke} have
recently shown that the parameter $\xi_t$ can also yield the
entanglement of formation (EOF) of the system, $E_F$,
\cite{Wooters}:
\begin{eqnarray}
E_F(\xi_t)&=&c_+(\xi_t)\log[c_+(\xi_t)]-c_-(\xi_t)\log[c_-(\xi_t)]
\;\;{\rm if}\; 0<\xi_t<1, \nonumber \\&=&0 \;\;{\rm otherwise,}
\label{EOF}
\end{eqnarray}
where the functions $c_{\pm}(z)$ are defined by
$c_{\pm}(z)=[(z)^{-1/2}\pm(z)^{1/2}]^2/4$. It is noteworthy that
EOF is a proper measure of the degree of entanglement between two
systems and is equal to the von-Neumann entropy if the compound
system remains in a pure state. However, unlike the von-Neumann
entropy, EOF still works for mixed states. In fact, Josse {\it et
al.} have recently measured the EOF of a pair of non-separable
light beams with this scheme \cite{Josse}. In the following
section, we shall make use of the two-mode entanglement parameter
$\xi_t$ to study how the two components of a BEC condensate
trapped in double-well are entangled.

\section{Two-mode entanglement in BECs}
To study two-mode entanglement in BECs, we first show that a
two-component BEC indeed forms a Gaussian system. Since
\begin{eqnarray}
\chi^{(w)}(\lambda_a,\lambda_b,t)
&=&\text{tr}\left\{{\rho}\text{exp} \left[i\sqrt{2}\Lambda^T
X(t)\right]\right\} \nonumber\\&=&\text{tr}\left\{{\rho}\text{exp}
\left[i\sqrt{2}\left(U(t)^T\Lambda\right)^T
X(t=0)\right]\right\}\,,
\end{eqnarray}
$\chi^{(w)}(\lambda_a,\lambda_b,t)$ can  be obtained from
$\chi^{(w)}(\lambda_a,\lambda_b,t=0)$ and we therefore consider
the characteristic function at $t=0$. For the initial state
$|0_a,0_b\rangle$, it is readily shown that
$\chi^{(w)}(\lambda_a,\lambda_b,t=0)$ is a Gaussian function  and
$M(t=0)$ is an identity matrix.
$\chi^{(w)}(\lambda_a,\lambda_b,t)$ at other times can simply be
obtained by the replacement $\Lambda \rightarrow U(t)\Lambda$.
Consequently, we  show that $\chi^{(w)}(\lambda_a,\lambda_b,t)$ is
a Gaussian function of $\Lambda$ and:
\begin{eqnarray}
\chi^{(w)}(\lambda_a,\lambda_b,t) &=&\text{exp}\left\{-\frac{1}{2}
\left[U^T(t)\Lambda\right]^T M(0) U^T(t)\Lambda \right\} \,,
\end{eqnarray}
which directly yields the covariance matrix $M(t)$:
\begin{eqnarray}
M(t) &=&U(t)M(0)U^{T}(t) \nonumber \\ &=&U(t)U^{T}(t).
\end{eqnarray}
Hence, the condensate constitutes a Gaussian system to which the
two-mode entanglement parameter applies.

After applying several local unitary transformations to $M(t)$, we
obtain the matrix $M_s(t)$ of our system, which reads:
\begin{displaymath} M_s(t)= \left( \begin{array}{clcr}
n & 0 &  c & 0 \\
0   &n&   0  & -c \\
c & 0 &  n & 0 \\
 0  &-c&   0  & n
\end{array} \right).
\end{displaymath}
Moreover, the matrix elements of the matrix $M_s$ are expressible
in terms of the variances of the physical quantities as follows:
\begin{eqnarray}
n&=&2\left\{\langle{q^2_a}(t)\rangle\langle{p^2_a}(t)\rangle
-[{\rm Re}\langle{q}_a(t){p}_a(t)\rangle]^2\right\}^{1/2}\nonumber \\
&=&2\left\{\langle{q^2_b}(t)\rangle\langle{p^2_b}(t)\rangle
-[{\rm Re}\langle{q}_b(t){p}_b(t)\rangle]^2\right\}^{1/2},\\
c&=&2\left[{\rm
Re}(\langle{{q}_a(t){p}_b(t)}\rangle\langle{{q}_b(t)
{p}_a}(t)\rangle -\langle{q_a(t){q}_b(t)}\rangle
\langle{p_a(t){p}_b(t)}\rangle)\right]^{1/2},
\end{eqnarray}
and the variances are given in the appendix for reference. It is
then obvious that our system is a symmetric one with
$n_1=n_2=m_1=m_2=n$ and hence $a_0=1$. Therefore, one can use the
corresponding two-mode entanglement parameter, which is given by
\begin{eqnarray}
\xi_t&=&n-c\,,
\end{eqnarray}
to evaluate the EOF of the system \cite{Giedke}.

\begin{figure}[ht]
\includegraphics[height=12cm]{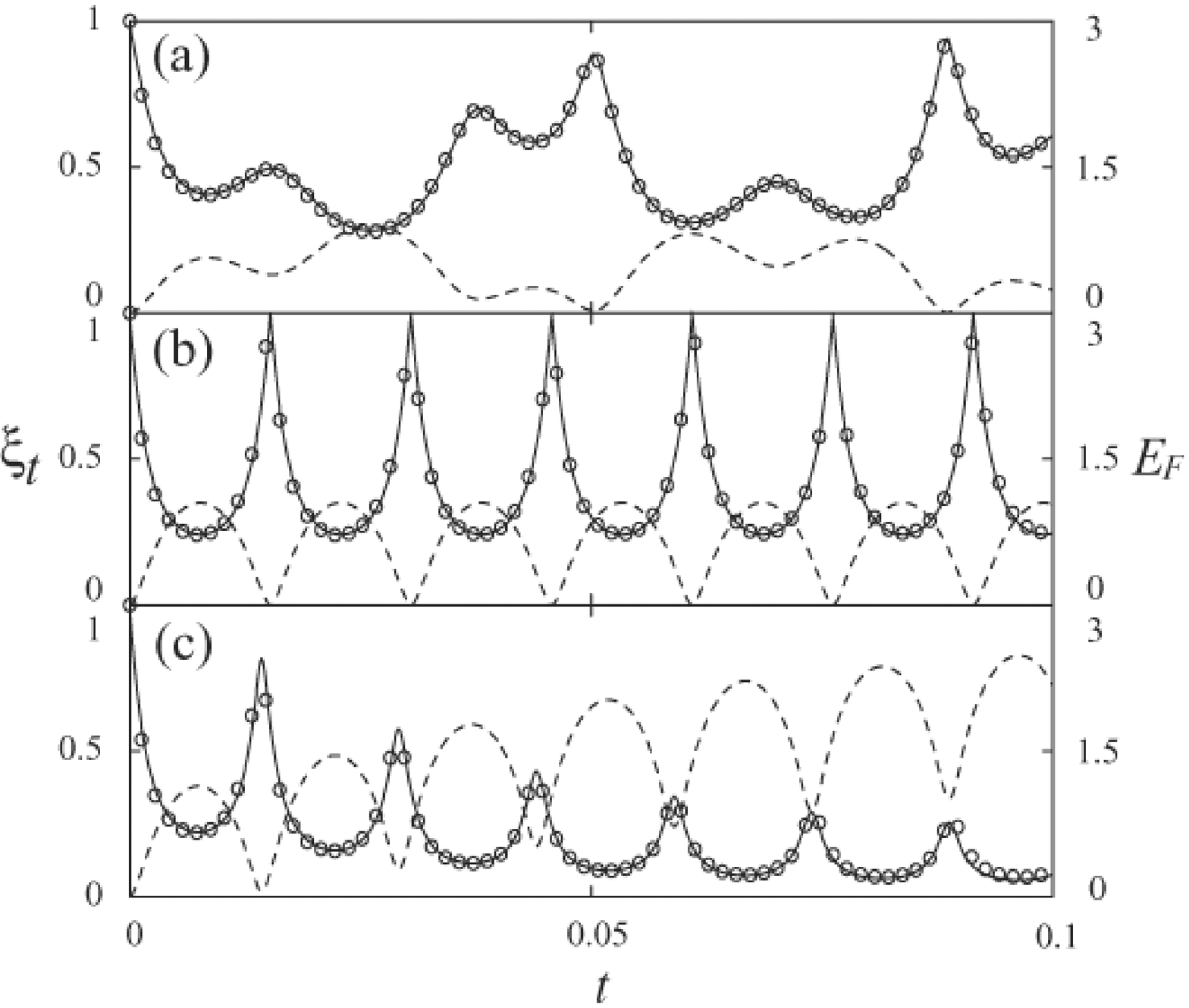}
\caption{\label{fig1} The time evolution of $\xi_t$ (solid line,
left scale) and the EOF $E_F$ (dashed line, right scale) are
shown for the symmetric case with $N_a=N_b=400$;
$\kappa_a=\kappa_b=1$; $\Omega_a=\Omega_b=50$; and (a)
$\kappa=0.50$; (b) $\kappa=1.00$; (c) $\kappa=1.12$. To gauge the
accuracy of the HPT, the exact numerical solution of $\xi_t$ is
shown by the empty circles.}
\end{figure}

Now we are ready to investigate the entanglement in our system
with the help of $\xi_t$ and $E_F(\xi_t)$. We first consider a
quasi-identical two-component condensate (e.g. a Rb-Rb
condensate), where $\kappa=\kappa_a=\kappa_b$ and
$\Omega_a=\Omega_b=\Omega$. The time evolution of $\xi_t$ for
such a condensate is shown in Fig.~\ref{fig1}, where
$N_a=N_b=400$, $\Omega_a=\Omega_b=50$, $\kappa_a=\kappa_b=1$ and
$\kappa=0.50$ (Fig.~\ref{fig1}(a)); $\kappa=1.00$
(Fig.~\ref{fig1}(b)); $\kappa=1.12$ (Fig.~\ref{fig1}(c)).
Hereafter we adopt suitable units in which $\kappa_b=1$ for
purpose of convenience. The solid line and empty circles are
respectively results obtained from the HPT and numerical
diagonalization of the original Hamiltonian, showing that the  HPT
indeed yields a good approximation in this regime. We also show
the time evolution of the EOF, which equals the von-Neumann
entropy in this case, by the dashed line in Fig.~\ref{fig1}. It is
obvious that our system is able to generate a substantial amount
of two-mode entanglement for most of the time. Besides, one can
see that there is a strong anti-correlation between these two
curves, which can be understood as $E_F(\xi_t)$ is a monotonically
decreasing function of $\xi_t$ for $0 \leq \xi_t \leq1$.

It is remarkable that the degree of entanglement depends crucially
on the inter-species interaction. Of course, it is obvious that
the EOF or the entropy is zero when the interspecies interaction
$\kappa$ vanishes. As shown in Fig.~\ref{fig1}(a), (b) and (c),
the entanglement parameter $\xi_t$ decreases while the EOF
increases with increasing $\kappa$.  Therefore, one can achieve
optimal squeezing by properly controlling the interaction
parameters. Besides, it is worthy of remark that in
Fig.~\ref{fig1}(c) $\kappa_c>\kappa>\kappa_e$. Therefore, the
system is still a stable one and the HPT remains valid. Physically
speaking, the availability of the two spatially separated modes in
fact stabilizes the condensate despite that $\kappa>\kappa_e$
\cite{phase-sep}.

We now switch our attention to asymmetric BECs consisting of two
components with different physical characteristics. In fact,
stable BECs of rubidium and potassium have recently been achieved
in experiments and it is also possible to control the strength of
interspecies interaction between the two components with a
magnetic field \cite{magint}. It is therefore deemed appropriate
to investigate how the entanglement in such condensates changes
with the interspecies interaction strength, $\kappa$. In the
following, we assume that the tunneling strengths and the
intraspecies interaction strengths of the two species are in the
ratios of 1:1.45 and 1:1.33 respectively, which are reasonable
estimates of experimental data for a Rb-K condensate \cite{walls}.

\begin{figure}
\includegraphics[height=12cm]{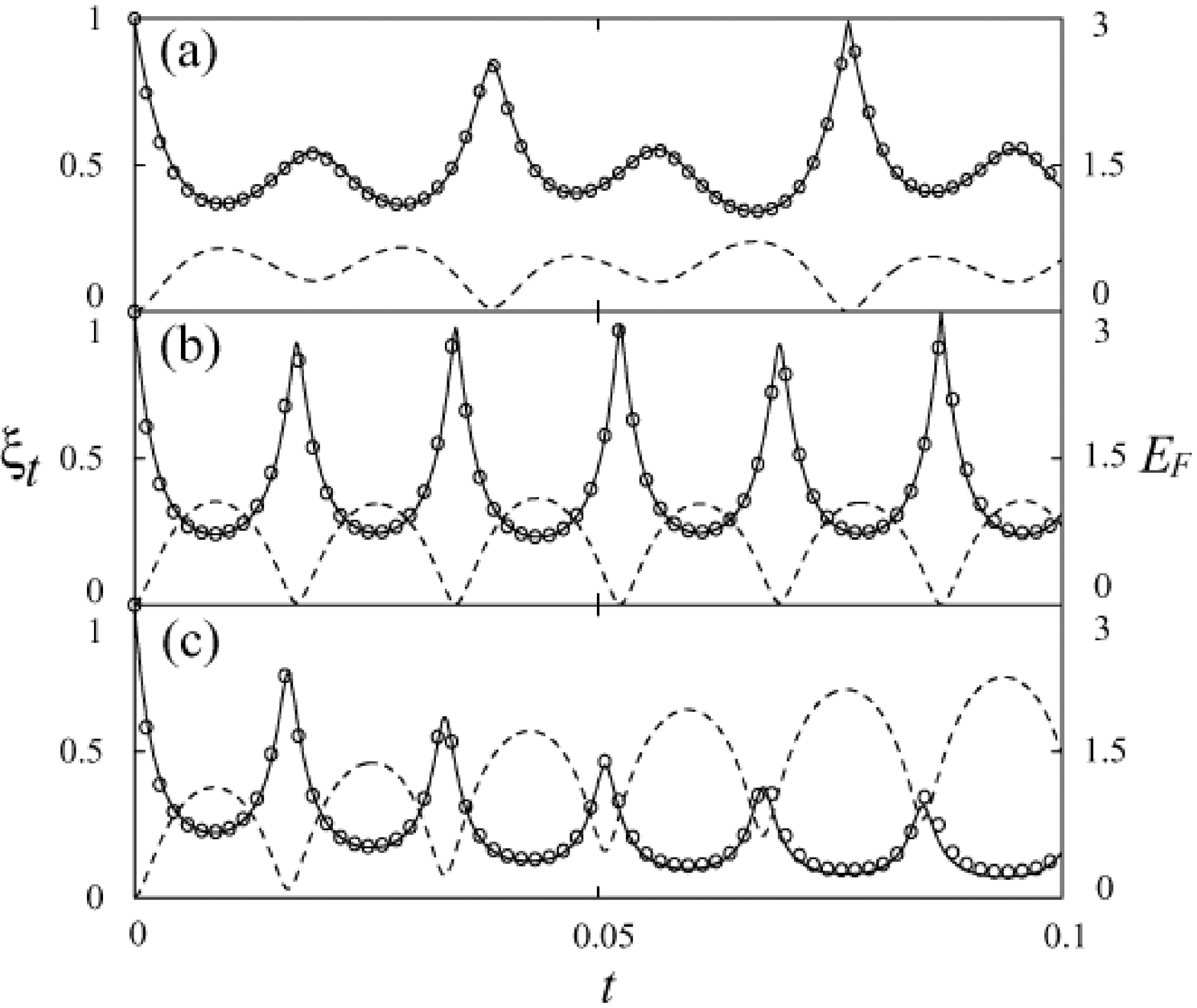}
\caption{\label{fig2} The time evolution of $\xi_t$ (solid line,
left scale) and the EOF $E_F$ (dashed line, right scale) are
shown for the asymmetric case with $N_a=N_b=400$;
$\kappa_a=0.75$, $\kappa_b=1$; $\Omega_a=34.5$, $\Omega_b=50$; and
(a) $\kappa=0.50$; (b)  $\kappa=0.875$; (c) $\kappa=0.965$. To
gauge the accuracy of the HPT, the exact numerical solution of
$\xi_t$ is shown by the empty circles.}
\end{figure}

The two-mode entanglement parameters for three asymmetric cases
with  $\kappa=0.5, 0.875, 0.965 $ are respectively shown in Fig.
\ref{fig2}(a), (b) and (c). Considering the interspecies
interaction $\kappa$ as an adjustable parameter, we find that a
smaller $\xi_t$ (i.e., higher entanglement) can be obtained as the
system becomes closer to the point of stability limit given by
(\ref{qsc}).  As shown in Fig.~\ref{fig2}(c), if $\kappa$ is
increased and approaches the stability limit  given by (\ref{qsc})
from below, the two-mode entanglement parameter (the EOF) can
attain much smaller (greater) values. Thus, the significance of
the strength of interspecies interaction in two-mode entanglement
generation is clearly demonstrated.

From the results illustrated in Figs.~1 and 2 it is manifest that
a substantial increase of entanglement can be achieved in the
vicinity of the stability limit (\ref{qsc}). In fact, this novel
feature can be understood heuristically as follows. In general,
position (momentum) squeezed states of a harmonic oscillator can
be produced by strengthening (weakening) its spring constant
\cite{Scully}. The two-component condensates can be viewed as a
coupled oscillators and at the critical point of stability the
eigenfrequenices  are
$(\Omega^2_a+\Omega^2_b+2\kappa_aj_a\Omega_a+2\kappa_bj_a\Omega_b)^{1/2}$
and zero. As one of these frequencies, $\omega_2 =0$, is markedly
different from those in the interaction free case, the effect of
squeezing in the position space and the momentum space is much
pronounced in the vicinity of the critical point.

\begin{figure}
\includegraphics[height=12cm]{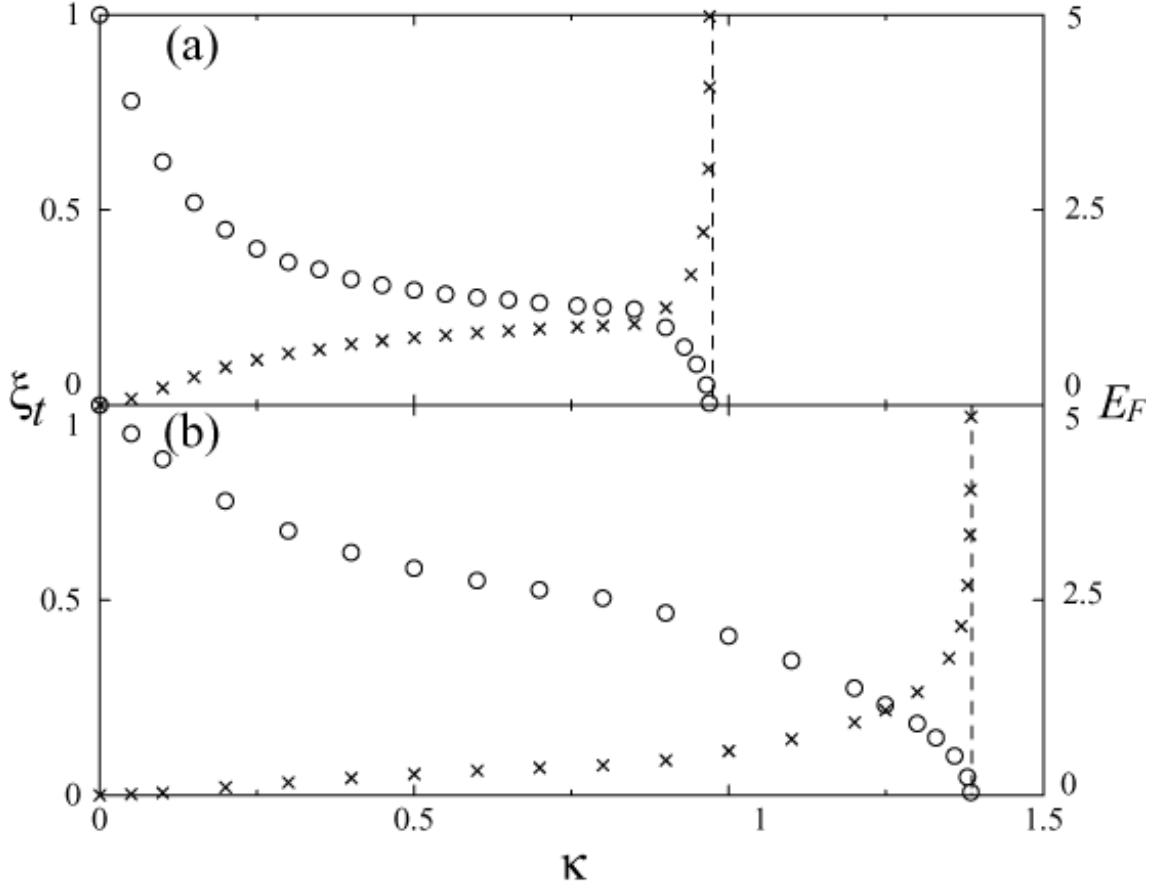}
\caption{\label{fig3} The maximal achievable values of  $\xi_t$
(empty circles, left scale) and  the EOF $E_F$ (crosses, right
scale) during time evolution are shown for different values of the
interspecies interaction strength $\kappa$. The condensate is an
asymmetric one with $N_a=N_b=400$; $\kappa_a=0.75$, $\kappa_b=1$;
and (a) $\Omega_a=34.5$, $\Omega_b=50$, $\kappa_c=0.969$; (b)
$\Omega_a=172.5$, $\Omega_b=250$, $\kappa_c=1.385$. The value of
$\kappa_c$ is shown by the vertical dashed line in each diagram.}
\end{figure}

To further elaborate this issue, we show the minimal value of
$\xi_t$ during the evolution of the coupled condensates and the
corresponding EOF as functions of the interspecies interaction
strength $\kappa$ in Fig \ref{fig3}, which explicitly confirms
that the degree of entanglement increases drastically as the
interspecies interaction $\kappa$ is close to the stability limit
given by (\ref{qsc}). In fact, both quantities change noticeably
once $\kappa > \kappa_e$. It is noteworthy that similar increase
in entanglement has previously been found in quantum phase
transition of a spin chain model \cite{Kitaev}.

\section{Discussion}
In the present paper the entanglement between the two components
of a BEC condensate trapped in a double-well is studied
analytically in the low excitation limit with the HPT, and its
accuracy is confirmed by comparison with the exact numerical
solution. As demonstrated in previous sections, the degree of
entanglement, gauged by the two-mode entanglement parameter,
depends strongly on the interspecies interaction that can be
varied with the current technology \cite{magint,Greiner}. We
expect that our work can be applied to study entanglement in
two-component condensates such as Rb-Rb and Rb-K mixtures.
Specifically, our result shows that the two components of the
condensate can remain in the tunneling phase and yet get strongly
entangled as long as $\kappa_c>\kappa>\kappa_e$.

To gain more physical insight from our result, we note that
$\hat{J}_{\alpha{x}}$ and $\hat{J}_{\alpha{y}}$ respectively
represent the population difference (measured by the operator
$\delta\hat{n}_\alpha \equiv
\hat{\alpha}_R^+\hat{\alpha}_R-\hat{\alpha}_L^+\hat{\alpha}_L$)
and the phase difference (measured by the relative phase operator
$\delta\hat{\phi}_\alpha$) of the $\alpha$-species condensate in
the two wells, where $\alpha=a,b$ \cite{Leggett}. Therefore, the
two-mode entanglement parameter, $\xi_t=n-c$,  where
\begin{eqnarray}
n&=&2\left\{\langle{\delta\hat{n}^2_a}(t)\rangle\langle{\delta\hat{\phi}^2_a}(t)\rangle
-[{\rm Re}\langle{\delta\hat{n}}_a(t){\delta\hat{\phi}}_a(t)\rangle]^2\right\}^{1/2}\nonumber \\
&=&2\left\{\langle{\delta\hat{n}^2_b}(t)\rangle\langle{\delta\hat{\phi}^2_b}(t)\rangle
-[{\rm Re}\langle{\delta\hat{n}}_b(t){\delta\hat{\phi}}_b(t)\rangle]^2\right\}^{1/2},\\
c&=&2\left[{\rm
Re}(\langle{\delta{\hat{n}}_a(t)\delta{\hat{\phi}}_b(t)}\rangle\langle{\delta{\hat{n}}_b(t)
\delta{\hat{\phi}}_a}(t)\rangle -\langle{
\delta\hat{n}_a(t){\delta\hat{n}}_b(t)}\rangle
\langle{\delta\hat{\phi}_a(t){\delta\hat{\phi}}_b(t)}\rangle)\right]^{1/2},
\end{eqnarray}
indeed measure the correlation of $\delta\hat{n}_\alpha$ and
$\delta\hat{\phi}_\alpha$. If the two components are entangled and
therefore $\xi_t<1$, the fluctuations in the population difference
and phase difference of the composite system are squeezed
accordingly.

On the other hand, in real experiments the temperature of the
condensate is not exactly zero \cite{Shin}. Therefore, it is worth
studying how the effect of finite temperature might affect the
entanglement that could be built up in the system during its
evolution when the initial state is a mixed state.  Specifically,
we consider an initial state that can be written as a product of
two thermally equilibrium states maintained at a common
temperature $T$, and the density matrix at $t=0$ is given by:
\begin{eqnarray}
\rho(0)&=&\rho_a\otimes\rho_b,
\end{eqnarray}
where for $\alpha=a,b$,
\begin{eqnarray}
\rho_{\alpha}&=&[1-{\rm exp}(-\Omega_{\alpha}/k_BT)]{\rm
exp}(-\Omega_{\alpha}\hat{a}^\dagger_{\alpha}\hat{a}_{\alpha}/k_BT).
\end{eqnarray}
It is a mixed stated and its two components are obviously
separable at $t=0$. The covariance matrix at $t=0$, $M(0)$, for
this initial state is:
\begin{displaymath} M(0)= \left( \begin{array}{clcr}
2\bar{n}_a+1 & 0 &  0 & 0 \\
0   &2\bar{n}_a+1&   0  & 0 \\
0 & 0 &  2\bar{n}_b+1 & 0 \\
 0  &0&   0  & 2\bar{n}_b+1
\end{array} \right),
\end{displaymath}
where, as usual, the mean excitation number $\bar{n}_\alpha$ is
$[{\rm exp}(\Omega_{\alpha}/k_BT)-1]^{-1}$. If the initial
temperature and the tunneling frequency are of order $10^{-8}~{\rm
K}$ and $1$~kHz respectively, which are typical values in current
experiment situations \cite{Greiner,Shin}, the mean excitation
number may reach order unity and can give rise to non-negligible
effect on the entanglement parameter.

\begin{figure}
\includegraphics[height=12cm]{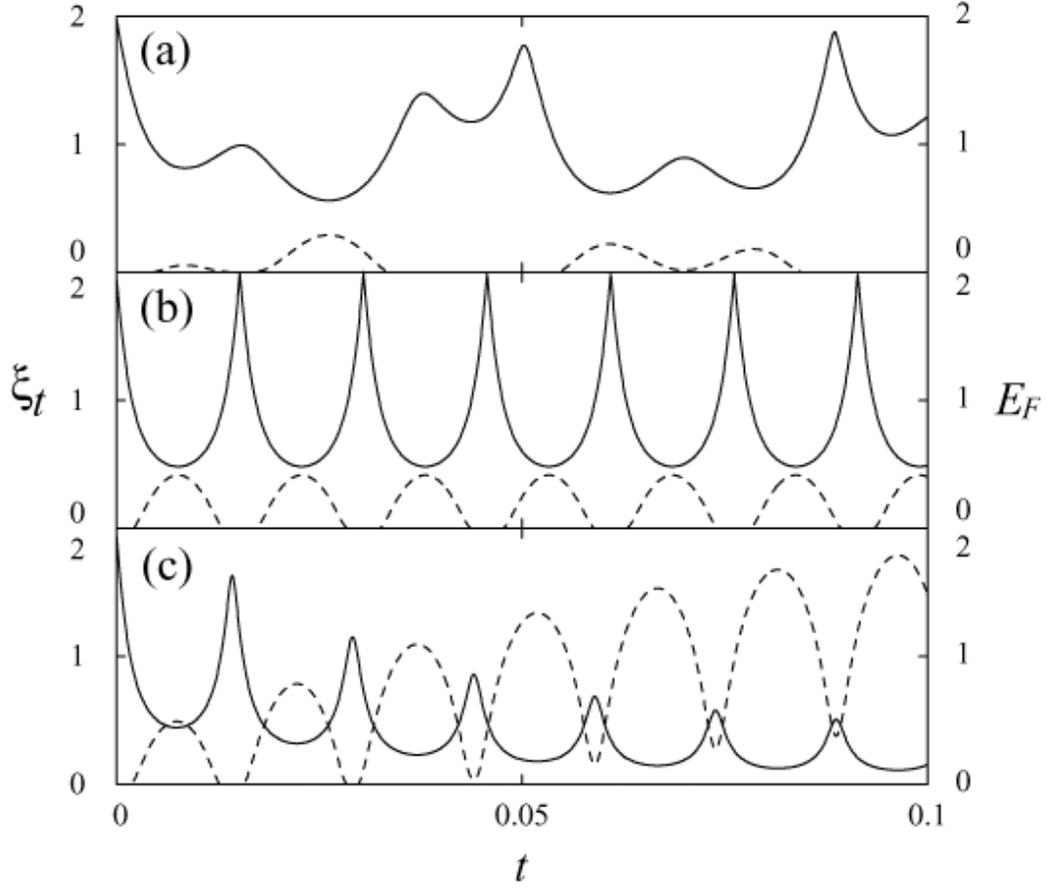}
\caption{\label{fig4} The time evolution of $\xi_t$ (solid line,
left scale) and the EOF $E_F$ (dashed line, right scale) for a
symmetric thermal BEC with $N_a=N_b=400$; $\kappa_a=\kappa_b=1$;
$\Omega_a=\Omega_b=50$; $\bar{n}_a=\bar{n}_b=0.5$; and (a)
$\kappa=0.50$; (b) $\kappa=1.00$; (c) $\kappa=1.12$.}
\end{figure}

It is well known that the Wigner characteristic function of a
harmonic oscillator in a thermal mixed state is still a Gaussian
function \cite{Barnett}. So, it is readily shown that the system
considered in our paper remains in a Gaussian state and one can
use the two-mode entanglement parameter to study the entanglement
between the condensates. Following the argument outlined
previously, one can show that the covariance matrix $M(t)$ at
$t>0$ is given by $U(t)M(0)U^{T}(t)$, from which the two-mode
entanglement parameter $\xi_t$ can be obtained accordingly. In
particular, for a symmetric two-component BEC with
$\Omega_a=\Omega_b$, the covariance matrix $M(t)$ is just
$U(t)U^{T}(t)$ multiplied by $2\bar{n}_a+1$ (or equivalently
$2\bar{n}_b+1$) and the system is again symmetric. The two-mode
entanglement parameter $\xi_t$ can then be used to determine
whether the system is entangled or not and to evaluate the
entanglement of formation $E_F(\xi_t)$ as well.  In
Fig.~$\ref{fig4}$ we show the two-mode entanglement parameter
$\xi_t$ and  the entanglement of formation $E_F$ as functions of
time for a symmetric two-component BEC with the mean excitation
number $\bar{n}_\alpha=0.5$. It is clearly manifested that a
substantial amount of entanglement can still be achievable when
$\kappa$ is sufficiently strong. Therefore, the existence of
finite thermal effects does not readily preclude the occurrence of
entanglement.

On the other hand, for an asymmetric two-component BEC with
$\Omega_a\neq\Omega_b$, the covariance matrix $M(t)$ is no longer
symmetric with respect to the two interacting components.
Therefore, despite that one can still make use of the parameter
$\xi_t$ to determine whether the system is entangled or not, it is
no longer possible to apply the method developed here to obtain
$E_F$ even if $\xi_t<1$ and other numerical schemes have to be
sought \cite{Wooters}.

In summary, although there might be some complications in
quantifying the degree of entanglement if the initial state of
the condensate is a mixed state, the two-mode entanglement
parameter studied here can still give a necessary and sufficient
condition of the separability of the two condensates.

\acknowledgments We thank CK~Law for helpful discussions and
comments on the manuscript. The work described in this paper was
partially supported by two grants from the Research Grants Council
of the Hong Kong Special Administrative Region, China (Project
Nos. 423701 and 401603).

\appendix
\section{Explicit form of the variances}
The explicit expressions of the variances are given here for
reference:
\begin{eqnarray}
\langle{\hat{q}^2_a}\rangle&=&\sum_{i=1,2}\Big\{{F_i}\cos2\omega_i{t}
+G_i\cos[\omega_1+(-1)^{i+1}\omega_2]t+H_i\Big\}\\
\langle{\hat{q}^2_b}\rangle&=&\frac{1}{\Omega_a}
\sum_{i=1,2}\Big\{\mu^2_i\Omega_a{F_i}\cos2\omega_i{t}
-\Omega_bG_i\cos[\omega_1+(-1)^{i+1}\omega_2]t+\mu^2_i\Omega_aH_i\Big\}\\
\langle{\hat{p}^2_a}\rangle&=&\frac{-1}{\Omega^{2}_a}
\sum_{i=1,2}\Big\{\omega^2_i{F_i}\cos2\omega_i{t}
+(-1)^{i+1}\omega_1\omega_2G_i\cos[\omega_1+(-1)^{i+1}\omega_2]t
\nonumber\\
&&+\omega^2_iH_i\Big\},\\
\langle{\hat{p}^2_b}\rangle&=&\frac{-1}{\Omega_a\Omega^{2}_b}
\sum_{i=1,2}\Big\{\mu^2_i\omega^2_i\Omega_a{F_i}
\cos2\omega_i{t}
+(-1)^i\omega_1\omega_2\Omega_bG_i\cos[\omega_1+(-1)^{i+1}\omega_2]t
\nonumber\\
&&+\mu^2_i\omega^2_i\Omega_aH_i\Big\},\\
{\rm Re}(\langle{\hat{q}_a\hat{q}_b}\rangle)&=&
\frac{1}{2}\sum_{i=1,2}\Big\{2\mu_i{F_i}\cos2\omega_i{t}
+(\mu_1+\mu_2)G_i\cos[\omega_1+(-1)^{i+1}\omega_2]t+2\mu_iH_i\Big\},\\
{\rm Re}(\langle{\hat{p}_a\hat{p}_b}\rangle)&=&
\frac{-1}{2\Omega_a\Omega_b}\sum_{i=1,2}\Big\{
2\mu_i\omega^2_i{F_i}\cos2\omega_i{t}\nonumber\\
&&+(-1)^{i+1}\omega_1\omega_2(\mu_1+\mu_2)G_i\cos[\omega_1+(-1)^{i+1}\omega_2]t
+2\mu_i\omega^2_iH_i\Big\},\\
{\rm Re}(\langle{\hat{q}_a\hat{p}_b}\rangle)&=&
\frac{-1}{2\Omega_b}\sum_{i=1,2}\Big\{2\mu_i\omega_i{F_i}{\sin}2\omega_i{t}
\nonumber\\
&&+[\mu_1\omega_1+(-1)^{i+1}\mu_2\omega_2]
G_i\sin[\omega_1+(-1)^{i+1}\omega_2]t\Big\},\\
{\rm Re}(\langle{\hat{q}_b\hat{p}_a}\rangle)&=&
\frac{-1}{2\Omega_a}\sum_{i=1,2}\Big\{2\mu_i\omega_i{F_i}\sin2\omega_i{t}
\nonumber\\
&&+[\mu_2\omega_1+(-1)^{i+1}\mu_1\omega_2]
G_i\sin[\omega_1+(-1)^{i+1}\omega_2]t\Big\},
\end{eqnarray}
where
\begin{eqnarray}
F_i&=&\frac{\tilde{\mu}^2_i}{4(\mu_1-\mu_2)^{2}}
\Bigg[1+\frac{1}{\tilde{\mu}^{2}_i}-\frac{\Omega^2_a}{\omega^{2}_i}
-\frac{\Omega^2_b}{\tilde{\mu}^{2}_i\omega^{2}_i}\Bigg],\\
G_i&=&\frac{1}{2(\mu_1-\mu_2)^{2}}
\Bigg[\frac{\Omega_b}{\Omega_a}-1
+(-1)^i\frac{\Omega_a\Omega_b}{\omega_1\omega_2}
+(-1)^{i+1}\frac{\Omega^2_b}{\omega_1\omega_2}\Bigg],\\
H_i&=&\frac{\tilde{\mu}^{2}_i}{4(\mu_1-\mu_2)^{2}}
\Bigg[1+\frac{1}{\tilde{\mu}^{2}_i}+\frac{\Omega_a}{\omega^{2}_i}
+\frac{\Omega^2_b}{\tilde{\mu}^{2}_i\omega^{2}_i}\Bigg],\\
\tilde{\mu}_j&=&\mu_1(1-\delta_{j1})+\mu_2(1-\delta_{j2}).
\end{eqnarray}

\newpage

\end{document}